\documentclass[12pt]{iopart}

\def\un{\leavevmode\hbox{\normalsize1\kern-4.6pt\large1}}

\newcommand{\rr}[2]{\mbox{$\langle \hat{c}_{#1}^{\dag} \hat{c}_{#2} \rangle$}}

\newcommand{\kk}[2]{\mbox{$\langle \hat{c}_{#1} \hat{c}_{#2} \rangle$}}
\newcommand{\ck}[2]{\mbox{$\langle \hat{c}_{#1}^{\dag} \hat{c}_{#2}^{\dag} \rangle$}}

\newcommand{\s}[4]{\mbox{$\langle \hat{c}_{#1}^{\dag} \hat{c}_{#2}^{\dag} \hat{c}_{#3} \hat{c}_{#4} \rangle$}}

\newcommand{\ksi}[4]{\mbox{$\langle \hat{c}_{#1}^{\dag} \hat{c}_{#2} \hat{c}_{#3} \hat{c}_{#4} \rangle$}}

\newcommand{\six}[6]{\mbox{$\langle \hat{c}_{#1}^{\dag} \hat{c}_{#2}^{\dag} \hat{c}_{#3}^{\dag} \hat{c}_{#4} \hat{c}_{#5} \hat{c}_{#6} \rangle$}}

\begin{document}

\title[Coupled condensate non-condensate quantum kinetics]{Self-consistent quantum kinetics of condensate and non-condensate via a coupled equations of motion formalism}
\author{N.P. Proukakis\dag}
\address{\dag\ Foundation for Research and Technology Hellas, Institute of Electronic Structure and Laser, P.O. Box 1527, Heraklion 711 10, Crete, Greece.}


\begin{abstract}

This paper extends an earlier quantum kinetics treatment for dilute, weakly-interacting, partially Bose-Einstein condensed gases, presented by the author elsewhere [J. Res. Natl. Inst. Stand. Technol. {\bf 101}, 457 (1996)], by consistently treating the dynamics of the uncondensed atoms to the same level of approximation as the condensed atoms. Our method is based on a hierarchy of coupled equations of motion for the condensate mean field and fluctuations around this mean field, truncated to second order in the (effective two-body) interatomic potential, and with suitable decoupling approximations for higher order correlations. By applying perturbation theory in the Hartree-Fock-Bogoliubov basis, we re-derive the quantum kinetic theory of Walser et al. [Phys. Rev. A {\bf 59}, 3878 (1999)], which further indicates the consistency of our treatment to the Kadanoff-Baym non-equilibrium Green's functions formalism for trapped gases.

\end{abstract}

\pacs{03.75.Fi, 05.70.Ln, 67.40.Db, 51.10.+y}

\maketitle

\section{Introduction}

The formulation of a consistent kinetic theory for the description of systems exhibiting Bose-Einstein condensation has been a very active area of research, since the very early days of the theoretical study of the behaviour of superfluid liquid helium \cite{Martin,Kadanoff}. Such pioneering work for homogeneous systems has led to the establishment of the essential theoretical framework, which has since been extended to the case of trapped condensates. Due to the extreme weakness of their interactions, in comparison to those of liquid helium, the recently formed Bose-Einstein condensates in alkali gases are ideal systems for testing the validity of such theories. 
There are at the moment various non-equilibrium approaches to the dynamics of dilute, weakly-interacting, trapped Bose-Einstein condensates. Nonetheless, despite recent progress in the field, there exists to date no uniquely accepted non-equilibrium theory for the coupled dynamics of condensate and thermal cloud. 
On the one hand, we find the theories of Stoof \cite{Stoof_FP} and Gardiner-Zoller et al. \cite{QKV}, based respectively on a non-perturbative Fokker-Planck equation for the non-equilibrium dynamics of the gas, and a master equation for the many-body density matrix. These have been compared to experiments \cite{MIT_Growth}, yielding very good agreement between them and with existing data for the issue of onset of condensation \cite{Stoof_Growth,QK_Growth}, and they further appear to be in reasonable qualitative agreement with earlier work of Kagan, Svistunov and Shlyapnikov \cite{QK_Growth,Kagan_QK}. Other types of kinetic theories (except that of Castin and Dum \cite{Castin_U1}) assume the existence of a mean field for condensed atoms and deal perturbatively with more complex correlations around the mean field, based on suitable decoupling approximations, as first discussed by Kirkpatrick and Dorfmann \cite{Kirkpatrick}. Such kinetic treatments have been recently discussed in the context of trapped gases by Proukakis and Burnett \cite{Prouk_NIST}, Zaremba-Nikuni-Griffin \cite{ZNG} and Walser-Williams-Cooper-Holland \cite{Walser_QK,Walser_Sim}. In a recent paper, Wachter et al. \cite{Wachter} have shown the equivalence of the non-equilibrium Green's function approach originally proposed by Kadanoff and Baym \cite{Kadanoff}, as applied to inhomogeneous systems by Imamovic-Tomasovic and Griffin \cite{Milena_Book,Milena_HFB}, to the second-order gapless kinetic theory of Walser et al. \cite{Walser_QK,Walser_Sim}. Central to this lies the fact that these theories yield the same second order damping rates, originally obtained in Beliaev's pioneering work \cite{Beliaev}, and later extended by Popov \cite{Popov}, and more recently by Fedichev and Shlyapnikov \cite{Gora}. 

Based on a hierarchy of coupled equations of motion formalism for the condensate mean field and fluctuations around it, the author and Burnett \cite{Prouk_NIST} have established an equation for condensate dynamics which extends beyond Hartree-Fock-Bogoliubov theory by the inclusion of triplet correlations; these correlations are related to the source field generating the condensate in the Green's functions approach, as discussed, for example, in  \cite{Martin,ZNG,Milena_Book}. Discussion of these equations has led to the ab initio introduction of an effective interaction in them \cite{Prouk_T_Matrix}, in a manner analogous to early diagrammatic work in the field \cite{Beliaev,Popov}. By focusing on static forms of these equations, Proukakis et al. \cite{Prouk_1D} have further discussed gapless many-body theories, extending also beyond the bare t-matrix, establishing at the same time relations with other approaches \cite{Bijlsma_Var,Shi}. Extending such work, Morgan \cite{Morgan} has formulated a gapless, number-conserving, self-consistent second order perturbation theory for the excitations of inhomogeneous condensates, which extends beyond the quasiparticle basis. Morgan has discussed at length the issues of gaplessness and infrared and ultraviolet divergences which have plagued the study of both homogeneous \cite{Beliaev,Popov,Gavoret,LHY,Gap} and inhomogeneous condensates from their very beginning.

In this paper, we extend the coupled equation of motion kinetic approach presented by the author elsewhere. In particular, in our preceeding work \cite{Prouk_NIST} we discussed how one formulates equations beyond the simplest HFB theory, by discussing the equations of motion of correlations of up to three single-particle operators. Despite containing all the essential dynamics for the condensate mean field, that treatment only dealt with correlations of four single-particle operators in their mean-field approximation, hence not being able to generate the dynamics of the non-condensate in a consistent manner. To achieve this, one must additionally look into the dynamic off-equilibrium contribution of four-operator correlations, which drive such correlations away from their respective equilibrium mean values due to interactions with the surrounding particles. Doing this above the transition point, the author \cite{Prouk_Thesis} successfully obtained the classical Boltzmann  equation, discussing at the same time how the actual interatomic potential becomes upgraded to an effective interaction. In this paper, we extend our earlier treatment by consistently treating, to second order in the effective (bare t-matrix) interaction, both the dynamics of the condensate and the non-condensate. This is based on the application of second order perturbation theory, starting from the usual Hartree-Fock-Bogoliubov (HFB), or quasiparticle basis, in much the same manner as the theory of Morgan \cite{Morgan}. Working in such a basis, we explicitly re-derive the second order theory of Walser et al. \cite{Walser_QK,Walser_Sim}, which is of a different nature, formulated essentially as a perturbative expansion of a many-body density operator around a self-consistently adjusted Gaussian and its correlations. This equivalence was anticipated, since both theories rely on the assumption of a few slowly-varying quantities, given in our case, by the HFB order parameters. By further noting that the theory of Walser et al. \cite{Walser_QK,Walser_Sim} was recently  shown \cite{Wachter} to be equivalent to existing theories based on Green's functions \cite{Milena_Book,Milena_HFB}, we hence additionally establish a connection with such theories. Furthermore, since our perturbative treatment deals with the same hamiltonian as the equilibrium theory of Morgan \cite{Morgan}, we further believe that this paper establishes an indirect link between  the static self-consistent approach of Morgan and the dynamic theory of Walser et al. We therefore consider this work as another step towards the formulation of a consistent theory for the study of both static and dynamic processes in finite temperature Bose-Einstein condensates.

The layout of this paper is as follows. Firstly we present our formalism, giving explicit expressions for the treatment of the hamiltonian of the system and the corresponding energy functional. We then introduce, in Sec. III, the generalised matrix notation for the condensate and quasiparticle propagators and self energies, in the usual manner. In Sec. IV, we give explicitly the first and second order evolution of these generalised propagators (in terms of the effective interatomic potential). Sec. V shows clearly the equivalence of our treatment to other kinetic theories (focusing on the recently proposed theory of Walser et al.), and we conclude this paper with some general remarks (Sec. VI). In the appendices, we explain how to deal with more complex normal and anomalous averages for both non-equilibrium contributions (via their respective equations of motion given in Appendix A) and their corresponding equilibrium values (for which we give the required decoupling approximations in Appendix B).

\section{Formalism}

We begin our treatment with the usual binary interaction Hamiltonian, 
\begin{equation}
\hat{H} = \sum_{rn} \Xi_{rn}  \hat{a}_{r}^{\dag} \hat{a}_{n} + \frac{1}{2}\sum_{rsmn} V_{rsmn} \hat{a}_{r}^{\dag} \hat{a}_{s}^{\dag} \hat{a}_{m} \hat{a}_{n} \label{Ham}
\end{equation}
which should incorporate most of the interesting physical processes occuring in dilute, weakly-interacting Bose-Einstein condensed gases.
Here  $\Xi = - (\hbar^{2} \nabla^{2})/(2m) + V_{trap}({\bf r})$ corresponds to the unperturbed hamiltonian in a harmonic trap,  and $V_{rsmn}$ represents the symmetrised form of the interaction potential between a pair of particles, defined by
$V_{rsmn}  = \frac{1}{2} \left\{ \langle rs| \hat{V} |  mn \rangle +   \langle rs| \hat{V} |  nm \rangle \right\}$. Here $|i \rangle = \psi_{i}({\bf r})$ denotes a single-particle eigenstate of the trap, and the single-particle operators $\hat{a}_{i}$ are related to the Bose field operator $\hat{\Psi}({\bf r},t)$ via
$\hat{\Psi}({\bf r},t) = \sum_{i} \psi_{i}({\bf r}) \hat{a}_{i}(t)$. Extending our earlier treatment regarding the ab initio introduction of an effective interaction \cite{Prouk_T_Matrix,Prouk_Thesis}, Morgan \cite{Morgan} has shown that, as long as high-lying states are adiabatically eliminated, and one is only interested in states up to a certain cut-off \cite{Popov}, the above hamiltonian can be equivalently written in terms of an effective re-summed two-body interaction. In fact, by numerically ensuring the independence of this effective interaction on the cut-off, this can be essentially replaced by the usual two-body t-matrix, often approximated by a local pseudopotential in three dimensions \cite{Huang,Prouk_JPhysB}. All our subsequent expressions will be given in terms of $V$, bearing in mind that this essentially corresponds to such a re-summed two-body effective interaction. Our formalism relies on the existence of symmetry breaking,  and hence we express the single-particle operators $\hat{a}_{i}$ as \cite{Blaizot}
\begin{equation}
\hat{a}_{i} = \langle \hat{a}_{i} \rangle + \left(\hat{a}_{i}- \langle \hat{a}_{i} \right) \rangle = z_{i} + \hat{c}_{i}
\end{equation}

One can formulate an infinite hierarchy of coupled equations of motion for the condensate mean field and fluctuations around this field. To make this problem tractable, certain approximations are required. Firstly, we note that for the very dilute gases we are dealing with, most of the interesting physics should be already apparent by their second order expressions (in the effective re-summed interatomic interaction), since such expressions contain both energy shifts and irreversible damping processes. Our theory further requires consistent decoupling approximations (discussed in Appendix B).
Due to the large number of atoms in existing condensates and the mean field potentials they generate, the single-particle eigenstates of the harmonic trap are not necessarily the best states for a perturbative expansion. Hence, starting from an HFB basis, we incorporate all remaining effects as a perturbation. To this aim, we write our hamiltonian as
\begin{equation}
\hat{H} = H_{0} + \hat{H}_{1}^{'} + \hat{H}_{2}^{'} + \hat{H}_{3}^{'} + \hat{H}_{4}^{'} 
\end{equation}
where 
\begin{equation}
H_{0} = \sum_{rn} \Xi_{rn} z_{r}^{*} z_{n} + \frac{1}{2} \sum_{rsmn} V_{rsmn} z_{r}^{*} z_{s}^{*} z_{m} z_{n}
\end{equation}
is the unpeturbed hamiltonian in a trap, additionally dressed by the condensate mean field, and the remaining operators are given by
\begin{eqnarray}
\fl \hat{H}_{1}^{'} & =  \sum_{rn} \Xi_{rn} \left[ \hat{c}_{r}^{\dag} z_{n} + z_{r}^{*} \hat{c}_{n} \right] + \sum_{rsmn} V_{rsmn} \left[ \hat{c}_{r}^{\dag} \left( z_{s}^{*} z_{m} + \rr{s}{m} \right) z_{n} + z_{r}^{*} \left( z_{s}^{*} z_{m} + \rr{s}{m} \right) \hat{c}_{n} \right] \nonumber \\
\fl & +   \frac{1}{2} \sum_{rsmn} V_{rsmn} \left[ \ck{r}{s} \hat{c}_{m} z_{n} + \hat{c}_{r}^{\dag} z_{s}^{*} \kk{m}{n} \right]
\end{eqnarray}

\begin{eqnarray}
\fl \hat{H}_{2}^{'} & =  \sum_{rn} \Xi_{rn}  \hat{c}_{r}^{\dag} \hat{c}_{n}  +  \sum_{rsmn} V_{rsmn} \left[ 2 \hat{c}_{r}^{\dag} \left( z_{s}^{*} z_{m} + \rr{s}{m} \right) \hat{c}_{n} -  \rr{r}{n} \rr{s}{m} \right] \nonumber \\
\fl  & +  \frac{1}{2} \sum_{rsmn} V_{rsmn} \left[ \hat{c}_{r}^{\dag} \hat{c}_{s}^{\dag} \left( z_{m} z_{n} + \kk{m}{n} \right) + \left( z_{r}^{*} z_{s}^{*} +\ck{r}{s} \right) \hat{c}_{m} \hat{c}_{n} - \ck{r}{s} \kk{m}{n} \right]
\end{eqnarray}

\begin{eqnarray}
\fl \hat{H}_{3}^{'} &  =   \sum_{rsmn} V_{rsmn} \left[ \left(  \hat{c}_{r}^{\dag} \hat{c}_{s}^{\dag} \hat{c}_{m} - 2 \hat{c}_{r}^{\dag} \rr{s}{m} - \ck{r}{s} \hat{c}_{m} \right) z_{n} \right] \nonumber \\
\fl & + \sum_{rsmn} V_{rsmn} \left[  z_{r}^{*} \left(  \hat{c}_{s}^{\dag} \hat{c}_{m} \hat{c}_{n} - 2 \rr{s}{m} \hat{c}_{n} - \hat{c}_{s}^{\dag} \kk{m}{n}  \right) \right]
\end{eqnarray}

\begin{eqnarray}
\fl \hat{H}_{4}^{'} & = \frac{1}{2} \sum_{rsmn} V_{rsmn} \left[  \hat{c}_{r}^{\dag} \hat{c}_{s}^{\dag} \hat{c}_{m} \hat{c}_{n} - 4 \hat{c}_{r}^{\dag} \rr{s}{m} \hat{c}_{n} - \hat{c}_{r}^{\dag} \hat{c}_{s}^{\dag} \kk{m}{n}  - \ck{r}{s} \hat{c}_{m} \hat{c}_{n} \right] \nonumber \\
\fl & + \frac{1}{2} \sum_{rsmn} V_{rsmn} \left[ 2 \rr{r}{n} \rr{s}{m} + \ck{r}{s} \kk{m}{n}  \right]
\end{eqnarray}
The preceeding operator-dependent hamiltonians ($\hat{H}_{i}^{'}$) have been denoted by a prime, indicating that these are not the usual hamiltonians one would obtain from equation (\ref{Ham}) upon writing $ \hat{a}_{i} = z_{i} + \hat{c}_{i} $, but they have been modified by the application of the following mean field approximations (see e.g. \cite{Griffin_Gap,Giorgini})
\begin{equation}
 \hat{c}_{r}^{\dag} \hat{c}_{s}^{\dag} \hat{c}_{m} \simeq \rr{r}{m} \hat{c}_{s}^{\dag} + \rr{s}{m} \hat{c}_{r}^{\dag} + \ck{r}{s} \hat{c}_{m}
\end{equation}
\begin{equation}
\hat{c}_{r}^{\dag} \hat{c}_{m} \hat{c}_{n} \simeq \rr{r}{m} \hat{c}_{n} + \rr{r}{n} \hat{c}_{m} + \hat{c}_{r}^{\dag} \kk{m}{n}
\end{equation}
\begin{eqnarray}
\fl \hat{c}_{r}^{\dag} \hat{c}_{s}^{\dag}  \hat{c}_{m} \hat{c}_{n} & \simeq   \hat{c}_{r}^{\dag} \rr{s}{m} \hat{c}_{n} +  \hat{c}_{r}^{\dag} \rr{s}{n} \hat{c}_{m} + \hat{c}_{s}^{\dag} \rr{r}{m} \hat{c}_{n} + \hat{c}_{s}^{\dag} \rr{r}{n} \hat{c}_{m} + \ck{r}{s} \hat{c}_{m} \hat{c}_{n} + \hat{c}_{r}^{\dag} \hat{c}_{s}^{\dag}  \kk{m}{n} \nonumber \\ 
\fl & -  \left( 2 \rr{r}{n} \rr{s}{m} + \ck{r}{s} \kk{m}{n}  \right)
\end{eqnarray}
Such mean field approximations are conventionally performed in HFB treatments aimed at reducing the  hamiltonian of the system to the usual quadratic form  $\left( \hat{H}_{0} + \hat{H}_{1}^{'} + \hat{H}_{2}^{'} \right)$, ignoring completely contributions beyond the mean field labelled above as $\left(\hat{H}_{3}^{'} + \hat{H}_{4}^{'} \right)$, since, by definition, $\langle \left(\hat{H}_{3}^{'} + \hat{H}_{4}^{'} \right) \rangle = 0$ when taking averages over HFB eigenstates. Such a hamiltonian can be diagonalised by a transformation to quasiparticle operators \cite{Bogoliubov}, which mix the shifted single-particle creation and annihilation operators $\hat{c}_{i}^{(\dag)}$. 
In such a theory, the effects of thermal atoms are treated only in an approximate manner, and this is known to lead to an unphysical gap in the spectrum of elementary excitations in the homogeneous limit \cite{Martin,Gap,Griffin_Gap}. This gap can be attributed to the inconsistent treatment of interactions between condensed and uncondensed components of the system and, more specifically, to the fact that the above approximation for three-particle operators is not physically justified  \cite{Prouk_T_Matrix,Morgan,Takano}. For consistency we note here that the approximation on four operators arises from an extension of Wick's theorem \cite{Blaizot} (which is rigorous for equilibrium averages) to the non-equilibrium case. 

To avoid such complications, we will hence work with an ``unperturbed'' hamiltonian consisting of the usual quadratic HFB hamiltonian $\left( H_{0} + \hat{H}_{1}^{'} + \hat{H}_{2}^{'} \right)$, but also keep the terms $\left( \hat{H}_{3}^{'} + \hat{H}_{4}^{'} \right)$ such that our hamiltonian is still exact, as also done by Morgan \cite{Morgan}. 
As we will show explicitly in this paper, this implies that all first-order contributions (assuming no steady state  anomalous correlations beyond the pair correlation) arise from the HFB hamiltonian $ \left( H_{0} + \hat{H}_{1}^{'} + \hat{H}_{2}^{'} \right)$, whereas the remaining parts of the hamiltonian $ \left( \hat{H}_{3}^{'} + \hat{H}_{4}^{'} \right) $  are {\em solely} responsible for {\em all} higher order collisional processes, damping rates, etc. In this paper, we will focus on the application of  the above hamiltonian to the dynamical domain, somewhat similarly to the approach of Zaremba-Nikuni-Griffin \cite{ZNG}. We start our treatment from a suitably generalised version of our earlier kinetic equations, and gradually build upon them, which leads to the explicit re-derivation of the kinetic theory of Walser et al. \cite{Walser_QK,Walser_Sim}.

\subsection{Some Definitions}

Applying perturbation theory on the HFB hamiltonian $ \left( H_{0} + \hat{H}_{1}^{'} + \hat{H}_{2}^{'} \right)$ amounts to assuming that the correlations of up to two single-particle operators  can be thought of as slowly-varying, whereas all other (higher order) correlations will be treated perturbatively (beyond their corresponding equilibrium contributions). We thus define the lowest order normal and anomalous averages of single-particle operators as
\begin{equation}
\rho_{ji} = \rr{i}{j} 
\end{equation}
\begin{equation}
\kappa_{ji}=\kk{i}{j} 
\end{equation}
These quantities, along with the condensate mean field $z_{i}$ form the set of order parameters of the HFB basis \cite{Blaizot}. This is analogous to the assumption made by Walser et al. \cite{Walser_QK} that such quantities form the set of relevant (master) variables for describing the evolution of the system on a coarse-grained timescale.
For a binary hamiltonian, as already noted in \cite{Prouk_Thesis}, a consistent treatment of the dynamic evolution of such quantities, to second order in the (effective) interatomic potential, requires an in-depth analysis of the evolution of all correlations, normal and anomalous, containing up to four single-particle shifted operators. As is well-known, higher order correlations containing an {\em even} number of shifted single-particle operators do not vanish at equilibrium, but actually acquire a steady state value consistent with Wick's theorem. This states that, at thermal equilibrium, such higher order averages can be decomposed into products of lower order averages, which in our case correspond to the slowly-varying HFB order parameters $z$, $\rho$ and $\kappa$. To proceed further with our coupled equation of motion methodology, we must hence treat such quantities in two steps \cite{Prouk_Thesis}: firstly we split them into their respective `equilibrium' and `non-equilibrium' components; subsequently, we decompose all equilibrium higher-order averages into products of HFB parameters, while at the same time perturbatively eliminating their rapidly evolving parts which drive them out of equilibrium. The non-equilibrium quantities we will be concerned with are defined by
\begin{equation}
\fl \sigma_{rsmn} = \s{r}{s}{m}{n} - \s{r}{s}{m}{n}^{(0)} = \s{r}{s}{m}{n} - \left( \rho_{mr} \rho_{ns} + \rho_{nr} \rho_{ms} + \kappa_{mn} \kappa_{rs}^{*} \right)
\end{equation}
and
\begin{equation}
\fl \xi_{rsmn}= \ksi{r}{s}{m}{n} - \ksi{r}{s}{m}{n}^{(0)} =  \ksi{r}{s}{m}{n} - \left( \rho_{sr}  \kappa_{mn} +  \rho_{mr}  \kappa_{ns}+  \rho_{nr}  \kappa_{ms} \right)
\end{equation}
This appearance of `equilibrium' values is not encountered by triplet correlations (or higher order {\em odd} single-particle operator averages). These are assumed to vanish identically at equilibrium (due to our choice of the HFB basis as a good basis for describing the equilibrium properties of the system) and can therefore be treated perturbatively. In this case, we define \cite{Prouk_NIST,Prouk_Thesis}
\begin{equation}
\lambda_{rmn}=\langle \hat{c}_{r}^{\dag} \hat{c}_{m} \hat{c}_{n} \rangle
\end{equation}
\begin{equation}
\gamma_{smn}=\langle \hat{c}_{s} \hat{c}_{m}  \hat{c}_{n} \rangle
\end{equation}

\subsection{Energy functional}

Having defined our notation, we can now explicitly write down the energy functional $\hat{E}$ of our system based on the hamiltonian (\ref{Ham}). Its {\em exact} form is given by
\begin{equation}
\hat{E} = \langle \hat{H} \rangle = \hat{E}^{(1)}_{HFB} + \hat{E}^{(2)} \label{EFTotal}
\end{equation}
Here we have defined the HFB energy functional in the usual manner \cite{Blaizot}
\begin{eqnarray}
\fl \hat{E}^{(1)}_{HFB} & \equiv  \langle \left( H_{0} + \hat{H}_{1}^{'} + \hat{H}_{2}^{'}
\right) \rangle = \sum_{rn} \Xi_{rn} \left( z_{r}^{*} z_{n} + \rho_{nr} \right) \nonumber \\
\fl & +  \frac{1}{2} \sum_{rsmn} V_{rsmn} \left[ z_{r}^{*} z_{s}^{*} z_{m} z_{n} + 4 \rho_{nr} z_{s}^{*} z_{m} + 2 \rho_{nr} \rho_{ms} + \kappa_{rs}^{*} z_{m} z_{n} + z_{r}^{*} z_{s}^{*} \kappa_{mn} + \kappa_{rs}^{*} \kappa_{mn} \right] \label{EF1}
\end{eqnarray}
and the additional `perturbative' beyond-HFB functional
\begin{eqnarray}
\hat{E}^{(2)} & \equiv & \langle \left(  \hat{H}_{3}^{'} + \hat{H}_{4}^{'}
\right) \rangle = \frac{1}{2} \sum_{rsmn} V_{rsmn} \left[ 2 \lambda_{mrs}^{*} z_{n} + 2 \lambda_{smn} z_{r}^{*} + \sigma_{rsmn} \right] \label{EF2}
\end{eqnarray}
Consideration of $ \hat{E}^{(1)}_{HFB} $ leads to the well-known (reversible) first order expression for the dynamics of the HFB order parameters. The second term,  $\hat{E}^{(2)} $ has no lowest order contribution in the HFB basis (since the HFB basis, by definition, does not allow for a non-vanishing steady state value for the triplets) and is hence conventionally ignored. 

In this paper we demonstrate that it is precisely this second contribution to the energy functional  which generates all second (and higher order) collisional contributions and should, thus, not be ignored \cite{Morgan}. In fact, we will show explicitly that the second order energy functional discussed by Walser et al.  \cite{Walser_Sim} is nothing but our functional $\left ( \hat{E}^{(1)}_{HFB} + \hat{E}^{(2)} \right) $ when treated to second order, thus justifying its particular form. To obtain the desired higher order contributions in a  perturbative expansion, we thus require the equations of motion of  $ \lambda $ and $\sigma $, as given in Appendix A. 

Making use of these expressions and re-arranging, we obtain for the second order expression of the energy functional
\begin{eqnarray}
\fl \hat{E}^{(2)} & = (-i) \sum_{rsmn} \sum_{pqlt} V_{rsmn} \tilde{V}_{pqlt} \nonumber \\
\fl & \times \left\{ \begin{array}{l}  \left[ \left( \rho_{mp} + \delta_{mp} \right) \left( \rho_{nq} + \delta_{nq} \right) \rho_{ts} \rho_{lr} - \rho_{mp} \rho_{nq} \left( \rho_{ts} + \delta_{ts} \right)   \left( \rho_{lr} + \delta_{lr} \right) \right] \\ 
+ 2  \left[ \left( \rho_{mp} + \delta_{mp} \right) \left( z_{q}^{*} z_{n} \right) \rho_{ts} \rho_{lr} - \rho_{mp} \left( z_{q}^{*} z_{n} \right) \left( \rho_{ts} + \delta_{ts} \right)  \left( \rho_{lr} + \delta_{lr} \right) \right] \\ 
+ 2 \left[ \left( \rho_{mp} + \delta_{mp} \right) \left( \rho_{nq} + \delta_{nq} \right) \left( z_{s}^{*} z_{t} \right) \rho_{lr} - \rho_{mp} \rho_{nq} \left( z_{s}^{*} z_{t} \right) \left( \rho_{lr} + \delta_{lr} \right) \right] \\  
+4 \left[ \left( \rho_{mp} + \delta_{mp} \right) \kappa_{nt} \kappa_{qs}^{*} \rho_{lr} - \rho_{mp} \kappa_{nt} \kappa_{qs}^{*}  \left( \rho_{lr} + \delta_{lr} \right)  \right] \\  
+4 \left[ \left( \rho_{mp} + \delta_{mp} \right) \kappa_{nt} \kappa_{qs}^{*}  \left( z_{r}^{*} z_{l} \right) - \rho_{mp} \kappa_{nt} \kappa_{qs}^{*}  \left( z_{r}^{*} z_{l} \right)  \right] \\  
+4 \left[ \left( z_{p}^{*} z_{m} \right) \kappa_{nt} \kappa_{qs}^{*} \rho_{lr} -  \left( z_{p}^{*} z_{m} \right)  \kappa_{nt} \kappa_{qs}^{*} \left( \rho_{lr} + \delta_{lr} \right)  \right] \\ 
+4 \left[ \left( \rho_{mp} + \delta_{mp} \right) \left( z_{n} z_{t} \right) \kappa_{qs}^{*} \rho_{lr} - \rho_{mp}  \left( z_{n} z_{t} \right)  \kappa_{qs}^{*} \left( \rho_{lr} + \delta_{lr} \right) \right] \\   
+4 \left[ \left( \rho_{mp} + \delta_{mp} \right) \kappa_{nt} \left( z_{q}^{*} z_{s}^{*} \right) \rho_{lr} - \rho_{mp} \kappa_{nt}  \left( z_{q}^{*} z_{s}^{*} \right)  \left( \rho_{lr} + \delta_{lr} \right)  \right]   
\end{array} \right\} \label{EF2Extended}
\end{eqnarray}
where $\delta_{ij}$ is the usual Cronecker delta.

We stress that the above expression only corresponds to the second order contribution of the energy functional $\hat{E}$, and not to the exact energy functional given by equations (\ref{EFTotal})-(\ref{EF2}). In particular, we note that this second order expression additionally contains higher order averages such as $\gamma$, $\lambda$, $\sigma$ and $\xi$ which play the crucial role of generating {\em all} higher order contributions to the energy functional (but  do not modify the system evolution to second order, just as $\hat{E}^{(2)}$ has no effect on {\em first} order expressions for condensate / non-condensate evolution).

In the above expression,  we have defined the approximately energy-conserving matrix element
\begin{equation}
\tilde{V}_{pqlt} = \int dt^{'} e^{- i \left( \tilde{\omega}_{l} +  \tilde{\omega}_{t} -  \tilde{\omega}_{p} -  \tilde{\omega}_{q} \right) (t - t^{'}) } =  \pi \delta(\Delta \tilde{\omega}) - i P \left( \frac{1}{ \Delta \tilde{\omega}} \right) \label{Matrix}
\end{equation}
where $ \Delta \tilde{\omega} = \left( \tilde{\omega}_{l} + \tilde{\omega}_{t} - \tilde{\omega}_{p} - \tilde{\omega}_{q} \right) $ and $ \tilde{\omega}_{i}$ denotes the energy of level $i$ dressed by HFB mean field potentials (i.e. expectation value in the quasiparticle hamiltonian $\left(H_{0} + \hat{H}_{1}^{'} + \hat{H}_{2}^{'} \right)$). 
We note that the particular structure of the energy-conserving indices in the above expression for the energy functional  (i.e. that they correspond precisely to the second matrix element $V$ of $\hat{E}_{2}$) is a direct consequence of the Markov approximation which has been  additionally imposed here (see section IV B). However, the Markov approximation  only affects the indices appearing in the quasiparticle energies of the exponential in the integrand, and not the general structure of the energy functional.

Apart from the first line of equation (\ref{EF2Extended}) containing the usual Boltzmann scattering rates for normal (uncondensed)  averages $\rho$, we find terms involving condensate populations $(z^{*}z)$, as well as condensed ($zz$) and uncondensed ($\kappa$) anomalous averages. The appealing feature of this energy functional (which, admittedly, is not apparent from its first order expression of equation (\ref{EF2})) is its explicit symmetry with respect to condensate and non-condensate contributions. For example, additionally to the usual `classical' Boltzmann rates (first line of equation (\ref{EF2Extended})), we obtain all similar contributions in which one of the normal uncondensed averages $\rho$ is replaced by $(z^{*}z)$. Even more appealing is the fact that this further holds when considering anomalous averages. Hence, in addition to terms $\sim \left[ (\rho + 1) \kappa \kappa^{*} \rho \right]$, $\hat{E}^{(2)}$ contains terms $\sim \left[ (\rho + 1) \kappa (zz)^{*} \rho \right]$, $\left[ (\rho + 1) (zz) \kappa \rho \right]$, $\left[ (\rho + 1) \kappa \kappa^{*} (z^{*}z) \right] $ and 
$ \left[ (z^{*}z) \kappa \kappa^{*} \rho \right]$, etc. We note that it is precisely the choice of the HFB basis for our unperturbed hamiltonian which prohibits, in the final expressions, the appearance of multiple products of normal or anomalous condensate averages beyond their simplest forms  $(z^{*}z)$, $(zz)$ or $(z^{*}z^{*})$. 
This simplification would not arise if we were working with a simpler unperturbed hamiltonian, such as the one describing bare trap eigenstates, the Gross-Pitaevskii basis (dressed only by the condensate mean field) or the Hartree-Fock basis (additionally dressed by the normal non-condensate average). Finally, as already pointed out by Walser et al. \cite{Walser_QK}, although the in and out rates for the normal uncondensed component differ due to the process of bosonic enhancement (modifying $\rho$ to $(\rho+1)$), the mean field $(z^{*}z)$ is never bosonically enhanced, and can thus be thought of as a classical field.

\section{Generalised matrix notation}

In order to present our formalism in the most general manner and establish a straighforward link with existing theories, we will henceforth work with generalised density matrices and hamiltonians, in a manner which  clearly distinguishes between first order HFB and higher order perturbative results. We hence define the generalised condensate matrix $R_{c}$ ($2n \times 1$) and the  generalised quasiparticle density matrix $R_{e}$ ($2n \times 2n$) by \cite{Blaizot}
\begin{eqnarray}
R_{c}= \left( \begin{array}{c}  z \\ z^{*} \end{array} \right) \hspace{2.0cm}  R_{e}= \left( \begin{array}{c}  \rho \\  \kappa^{*} \end{array} \begin{array}{c} \kappa \\  (\rho^{*} + {\bf \un}) \end{array} \right)
\end{eqnarray}
where $\un$ is the unity matrix of the n-dimensional Fock space.
These quantities can be thought of as the propagators for the condensed and uncondensed parts of the system, and are analogous to the corresponding generalised propagators conventionally employed in Green's functions formulations \cite{Martin,Kadanoff,Milena_Book,Milena_HFB,Shi,Griffin_Gap,Milena_QP}. We note that, although $R_{e}$ includes all of the effects of a quasiparticle basis, these are explicitly given in terms of correlations of {\em single-particle} operators, with an explicit transformation to quasiparticle basis being beyond the scope of this paper (see e.g. \cite{Milena_QP,Wachter_Thesis}).

Correspondingly, we define the generalised condensate and quasiparticle hamiltonians as
\begin{eqnarray}
H_{c} = \left( \begin{array}{c} h^{(c)} \\  - (\Delta^{(c)})^{*} \end{array} \begin{array}{c}    \Delta^{(c)} \\ -  (h^{(c)})^{*}   \end{array} \right)
\hspace{2.0cm}
H_{e} = \left( \begin{array}{c} h \\ - \Delta^{*} \end{array} \begin{array}{c}    \Delta \\ -  h^{*}   \end{array} \right)
\end{eqnarray}
which can be thought of as the generalised self-energies for the above Green's functions.

These generalised  HFB matrix hamiltonians are already known from variational approaches, in which they can be readily obtained by variation of the energy functional $\hat{E}_{HFB}^{(1)}$ with respect to the corresponding HFB propagators. In particular, one finds \cite{Blaizot}
\begin{eqnarray}
\begin{array}{c}  {h}_{ij} \equiv \frac{\partial \hat{E}_{HFB}^{(1)}}{\partial \rho_{ji}(t)} =  {h}^{c}_{ij} +  \sum_{kl} V_{iklj} \left( z_{k}^{*} z_{l} \right) = \langle i| \hat{\Xi} | j \rangle + 2 \sum_{kl} V_{iklj} \left[z_{k}^{*} z_{l}  + \rho_{lk} \right] \\
\Delta_{ij}  \equiv  \frac{\partial \hat{E}_{HFB}^{(1)}}{\partial \kappa_{ji}^{*}(t)} =  \Delta^{c}_{ij} + \sum_{kl} V_{ijkl} \left( z_{k} z_{l} \right) =  \sum_{kl} V_{ijkl} \left[ z_{k} z_{l} + \kappa_{kl}  \right] \end{array}
\end{eqnarray}

For consistency, and ease of subsequent comparison to the expressions of Walser et al. \cite{Walser_Sim}, in this paper we have defined the generalised hamiltonians $H_{c}$ and $H_{e}$ in a slightly different manner from those of our earlier treatment \cite{Prouk_NIST,Prouk_Thesis}. This difference is only a matter of notation and arises from how the Pauli matrix $\sigma^{(3)} = \left( \begin{array}{l} \un \\ 0 \end{array} \begin{array}{c} 0 \\ -  \un  \end{array} \right)$ is incorporated in our equations\footnote{We also note that, despite the different notation, our previous presentation contained a minor error, in that the left hand sides of equations (32) and (2.32) of \cite{Prouk_NIST} and \cite{Prouk_Thesis} respectively, should contain an additional factor of $\eta=\sigma^{(3)}$).}. We also point out the useful identities $h_{ij}^{*} = h_{ji}$, $\Delta_{ij} = \Delta_{ji}$, $\rho_{ji}=\rho_{ij}^{*}$ and $\kappa_{kj}=\kappa_{jk}$, whereas all higher order correlations are symmetric with respect to the interchange of indices labelling operators of the same `type' (i.e. creation $\hat{c}^{\dag}$ or annihilation $\hat{c}$).

\section{Coupled evolution of condensate and quasiparticle propagators}

In this section  we discuss the dynamic evolution of the condensate and quasiparticle propagators defined above, by employing the Heisenberg equation of motion (setting $\hbar=1$ for simplicity)
\begin{equation}
i \frac{d}{dt} \langle \hat{O} \rangle  =  \langle \left[ \hat{O}, \hat{H} \right] \rangle
\end{equation}
for the mean value of a general operator $ \hat{O} $.

\subsection{Exact first order evolution}

To first order in the effective potential, the equations of motion for the condensed and uncondensed propagators acquire the following {\em exact} form \cite{Prouk_NIST,Prouk_Thesis}

\begin{eqnarray}
i \frac{d R_{c}}{dt} =  H_{c} R_{c} + J \label{CondProp}
\end{eqnarray} 

\begin{equation}
i \frac{d R_{e}}{dt} =  \left( H R - R H^{\dag} \right) + K \label{QPProp}
\end{equation}

In each of the above equations, the first terms correspond to the HFB contributions, arising entirely from the hamiltonian $\left( H_{0} + \hat{H}_{1}^{'} + \hat{H}_{2}^{'} \right) $. The static form of such equations can be readily derived variationally from $\hat{E}_{HFB}^{(1)}$ \cite{Blaizot}. Our approach extends beyond such treatments, in that it further generates the `non-equilibrium' matrices $J$ and $K$ for condensate and non-condensate evolution. These terms, arising solely from $\hat{E}^{(2)}$, depend on averages of higher order than the HFB parameters, and are thus conventionally set to zero at equilibrium. In this work, these matrices are treated perturbatively, thus generating terms of higher order in the potential, which can still depend on the non-vanishing equilibrium quantities $z$, $\rho$ and $\kappa$. This shows that such contributions, and hence the beyond-HFB energy functional $\hat{E}^{(2)}$ cannot be ignored even in static treatments, a point discussed clearly in the related work of Morgan \cite{Morgan}. To further establish a connection to kinetic treatments involving a damped nonlinear Schr\"{o}dinger equation for the condensate mean field at finite temperatures, we note that it is precisely the presence of such additional contributions (labelled here by the kinetics matrix $J$) which lead to the damping term. In fact, by carefully identifying the physical origin of such a damping term (in a manner somewhat analogous to the present treatment), Zaremba et al. \cite{ZNG} and Bijlsma et al. \cite{Stoof_Growth} have respectively performed detailed studies of the hydrodynamic regime and the onset of condensation. It further appears that the kinetic matrix $J$ is also taken into account in treatments based on linearised equations of motion, either explicitly \cite{Rusch}, or implicitly \cite{Giorgini}.

Coming back to our formalism, the {\em exact} form of these kinetic matrices is given by
\begin{eqnarray}
J = \left( \begin{array}{c} L  \\ - L^{*} \end{array}  \right)
\end{eqnarray}

\begin{eqnarray}
K = \left( \begin{array}{c} (M -  \widetilde{M}^{*}) \\ - (N +  \widetilde{N})^{*} \end{array} \begin{array}{c}  (N + \widetilde{N})  \\  - (M -  \widetilde{M}^{*})^{*}  \end{array} \right)
\end{eqnarray}
where
\begin{equation}
L_{ji} =  \sum_{smn} V_{jsmn}\lambda_{smn}
\end{equation}

\begin{equation}
M_{ji} = \sum_{smn} V_{jsmn} [\sigma_{ismn} + \lambda_{imn} z_{s}^{*} + \lambda_{mis}^{*} z_{n}+ \lambda_{nis}^{*} z_{m}  ]
\end{equation}

\begin{equation}
N_{ji} = \sum_{smn} V_{jsmn} [\xi_{ismn} + \lambda_{sim} z_{n} + \lambda_{sin} z_{m}+ \gamma_{imn} z_{s}^{*}  ]
\end{equation}
and $  \widetilde{N}$ represents the transpose of matrix $N$.
Note that the above definitions follow the corresponding definitions of \cite{Prouk_NIST,Prouk_Thesis}, which they generalise by additional inclusion of quartic terms $\sigma$ and $\xi$ required for the consistent treatment of the non-condensate kinetics.

\subsection{Second order collisional integrals}

To proceed further and derive all relevant second order collisional integrals, we merely need to derive the corresponding equations of motion for the above kinetic matrices $J$ and $K$. This requires the evolution of higher order normal and anomalous single-particle operator correlations, as given in Appendix A, and the application of consistent decoupling approximations, given in Appendix B, for truncating the infinitely coupled equation of motion hierarchy. Our choice to apply perturbation theory starting from an HFB unperturbed basis, implies that the only effect of the hamiltonian $\left(H_{0} + \hat{H}_{1}^{'} + \hat{H}_{2}^{'} \right)$, beyond yielding the first order evolution, is to define the renormalised unperturbed eigenenergies $\tilde{\omega}_{i}$ which appear in all higher-order expressions. In particular, we note that these eigenenergies dressed by HFB mean fields are implicit in  the second-order collisional evolution of condensate / non-condensate propagators via the approximately energy-conserving matrix element $\tilde{V}_{pqlt}$ of equation (\ref{Matrix}). To keep our subsequent notation compact, we henceforth suppress the free evolution of matrices $J$ and $K$ (see also Appendix A). Moreover, in giving their respective {\em lowest order} collisional evolution, we additionally impose the Markov approximation on the slowly-evolving HFB order parameters $z$, $\rho$ and $\kappa$ appearing on the right hand sides of equations (\ref{CondKin})-(\ref{QPKin}). This relies on the assumption that such quantities evolve significantly only  over many complete collisional events, and thus decay much slower than all other interparticle correlations. We expect this to be a valid assumption for the dilute, weakly-interacting systems under consideration. (A detailed discussion of non-Markovian kinetic treatments which also deal explicitly with initial correlations in a system can be found in \cite{Morozov}.) Within the above simplifications, we thus obtain 
\begin{eqnarray}
i  \frac{dJ}{dt} = \left( \begin{array}{c} \Gamma_{z} \\  \left[ \Gamma_{z^{*}} \right]^{*}  \end{array} \begin{array}{c} \Gamma_{z^{*}}  \\ \left[ \Gamma_{z} \right]^{*}  \end{array} \right)  \left( \begin{array}{c} z \\ z^{*} \end{array} \right) \label{CondKin}
\end{eqnarray}
\begin{eqnarray}
\fl i  \frac{dK}{dt} = \left( \begin{array}{c} \left[ \Gamma_{\rho} \rho + \Gamma_{\kappa} \kappa^{*} + I_{\rho} \right] \\ \left[  \Gamma_{\rho} \kappa +  \Gamma_{\kappa} \left( \rho + \un \right)^{*} + I_{\kappa} \right]^{*}  \end{array} \begin{array}{c}  \left[  \Gamma_{\rho} \kappa +  \Gamma_{\kappa} \left( \rho + \un \right)^{*} + I_{\kappa} \right]  \\  \left[ \Gamma_{\rho} \rho + \Gamma_{\kappa} \kappa^{*} + I_{\rho} \right]^{*}  \end{array} \right) + h.c. \label{QPKin}
\end{eqnarray}
where $h.c.$ denotes the hermitian conjugate and we have used a compact matrix notation, writing, for example, $\left[ \left( \Gamma_{\rho} \right) \rho \right]_{ji} = \sum_{l} \left( \Gamma_{\rho} \right)_{jl} \rho_{li}$, and similarly for the other contributions.

We find that the rates $\Gamma_{\rho}$ and $\Gamma_{\kappa}$ appearing in the above equation can be calculated variationally from  the explicit second order expression for the energy functional of equation (\ref{EF2Extended}), as
\begin{equation}
\left( \Gamma_{\rho} \right)_{rl} = i \left( \frac{ \partial \hat{E}^{(2)} }{ \partial \rho_{lr} } \right)
\end{equation}
\begin{equation}
\left( \Gamma_{\kappa} \right)_{sq} = i \left( \frac{ \partial \hat{E}^{(2)} }{ \partial \kappa_{qs}^{*} } \right) 
\end{equation}
It is also interesting to note that 
\begin{equation}
\left( I_{\rho} \right)_{rl} = i \left( \frac{ \partial \hat{E}^{(2)} }{ \partial \left( \rho_{lr} + \delta_{lr} \right) } \right) 
\end{equation}
whereas we have not been able to obtain a correspondingly simple expression for $I_{\kappa}$. 

For the condensate rates $\Gamma_{z^{(*)}}$ we find
\begin{equation}
i \left( \frac{ \partial \hat{E}^{(2)} }{ \partial z_{i}^{*}  } \right) = \sum_{l} \left( \Gamma_{z} \right)_{il} z_{l}   + \sum_{q} \left( \Gamma_{z^{*}} \right)_{iq} z_{q}^{*} = - \left[ \frac{d z_{i}}{dt} \right]_{V^{2}}
\end{equation}
where the subscript $V^{2}$ has been used to indicate second-order contributions to the evolution of the condensate mean field. Unlike the case of the condensate propagator, the presence of the non-vanishing matrices $I_{\rho}$, $I_{\kappa}$ in equation (\ref{QPKin}) indicate that, in this case, one may not be able to define variationally a simple generalised hamiltonian including {\em all} rates (as was done earlier for the first order expressions).

Below we define explicitly the form of all  ($n \times n$) matrices appearing in the above equations, in terms of their respective elements
\begin{eqnarray}
\fl \left( \Gamma_{z} \right)_{il} = 2 \sum_{smn} \sum_{pqlt} V_{jsmn} \tilde{V}_{pqlt} \left\{ \begin{array}{l} \left[ \left( \rho_{mp} + \delta_{mp} \right) \left( \rho_{nq} + \delta_{nq} \right) \rho_{ts} - \rho_{mp} \rho_{nq} \left( \rho_{ts} + \delta_{ts} \right) \right] \\ + 2  \left[ \left( \rho_{mp} + \delta_{mp} \right) \kappa_{nt} \kappa_{qs}^{*} -  \rho_{mp} \kappa_{nt} \kappa_{qs}^{*} \right]  \end{array} \right\} 
\end{eqnarray}
\begin{eqnarray}
\fl \left( \Gamma_{z^{*}} \right)_{iq} = 2 \sum_{smn} \sum_{pqlt} V_{jsmn} \tilde{V}_{pqlt} \left\{ 2 \left[ \left( \rho_{mp} + \delta_{mp} \right) \kappa_{nt} \rho_{ls} -  \rho_{mp} \kappa_{nt} \left( \rho_{ls} + \delta_{ls} \right) \right] \right\} \label{Gammaz}
\end{eqnarray}
\begin{eqnarray}
\fl \left( \Gamma_{\rho} \right)_{jl} = 2 \sum_{smn} \sum_{pqlt} V_{jsmn} \tilde{V}_{pqlt} \left\{ \begin{array}{l}  \left[ \left( \rho_{mp} + \delta_{mp} \right) \left( \rho_{nq} + \delta_{nq} \right) \rho_{ts} - \rho_{mp} \rho_{nq} \left( \rho_{ts} + \delta_{ts} \right) \right] \\ + 2  \left[ \left( \rho_{mp} + \delta_{mp} \right) \left( z_{q}^{*} z_{n} \right) \rho_{ts} - \rho_{mp} \left( z_{q}^{*} z_{n} \right) \left( \rho_{ts} + \delta_{ts} \right) \right] \\ + \left[ \left( \rho_{mp} + \delta_{mp} \right) \left( \rho_{nq} + \delta_{nq} \right) \left( z_{s}^{*} z_{t} \right) - \rho_{mp} \rho_{nq} \left( z_{s}^{*} z_{t} \right) \right] \\  +2 \left[ \left( \rho_{mp} + \delta_{mp} \right) \kappa_{nt} \kappa_{qs}^{*} - \rho_{mp} \kappa_{nt} \kappa_{qs}^{*} \right] \\  +2 \left[ \left( z_{p}^{*} z_{m} \right) \kappa_{nt} \kappa_{qs}^{*} -  \left( z_{p}^{*} z_{m} \right)  \kappa_{nt} \kappa_{qs}^{*} \right] \\ +2 \left[ \left( \rho_{mp} + \delta_{mp} \right) \left( z_{n} z_{t} \right) \kappa_{qs}^{*} - \rho_{mp}  \left( z_{n} z_{t} \right)  \kappa_{qs}^{*} \right] \\   +2 \left[ \left( \rho_{mp} + \delta_{mp} \right) \kappa_{nt} \left( z_{q}^{*} z_{s}^{*} \right) - \rho_{mp} \kappa_{nt}  \left( z_{q}^{*} z_{s}^{*} \right)   \right]   \end{array} \right\}
\end{eqnarray}

\begin{eqnarray}
\fl \left( \Gamma_{\kappa} \right)_{jq} = 2 \sum_{smn} \sum_{pqlt} V_{jsmn} \tilde{V}_{pqlt} \left\{ \begin{array}{l} 2  \left[ \left( \rho_{mp} + \delta_{mp} \right) \kappa_{nt} \rho_{ls} - \rho_{mp} \kappa_{nt}  \left( \rho_{ls} + \delta_{ls} \right) \right] \\ 
 + 2  \left[ \left( \rho_{mp} + \delta_{mp} \right) \kappa_{nt} \left( z_{s}^{*} z_{l} \right) - \rho_{mp} \kappa_{nt}  \left( z_{s}^{*} z_{l} \right)  \right] \\
+ 2  \left[ \left( z_{p}^{*} z_{m} \right) \kappa_{nt} \rho_{ls} -  \left( z_{p}^{*} z_{m} \right) \kappa_{nt}  \left( \rho_{ls} + \delta_{ls} \right) \right] \\
+ 2  \left[ \left( \rho_{mp} + \delta_{mp} \right) \left( z_{n} z_{t} \right) \rho_{ls} - \rho_{mp} \left( z_{n} z_{t} \right) \left( \rho_{ls} + \delta_{ls} \right)   \right]    \end{array} \right\}
\end{eqnarray}
\begin{eqnarray}
\fl \left( I_{\rho} \right)_{ji} & = - 2 \sum_{smn} \sum_{pqlt} V_{jsmn} \tilde{V}_{pqlt}  \left( \delta_{li} \right) \nonumber \\
\fl & \times \left\{ \begin{array}{l} \rho_{mp}  \rho_{nq}  \left(  \rho_{ts} + \delta_{ts} \right) + 2  \left( z_{p}^{*} z_{m} \right) \rho_{nq}  \left(  \rho_{ts} + \delta_{ts} \right) +  \rho_{mp}  \rho_{nq} \left( z_{s}^{*} z_{t} \right) \\  + 2 \rho_{mp} \kappa_{nt} \kappa_{qs}^{*} + 2  \left( z_{p}^{*} z_{m} \right)  \kappa_{nt} \kappa_{qs}^{*}  + 2 \rho_{mp}  \left( z_{n} z_{t} \right) \kappa_{qs}^{*}  + 2 \rho_{mp} \kappa_{nt}  \left( z_{q}^{*} z_{s}^{*} \right) \end{array} \right\}
\end{eqnarray}
\begin{eqnarray}
\fl \left( I_{\kappa} \right)_{ji} & = 2 \sum_{smn} \sum_{pqlt} V_{jsmn} \tilde{V}_{pqlt} \left( \delta_{li} \right) \nonumber \\
\fl & \times \left\{ \begin{array}{l}  2 \rho_{mp} \kappa_{nt} \left(  \rho_{ls} + \delta_{ls} \right) + 2 \rho_{mp} \kappa_{nt} \left(  z_{s}^{*} z_{l} \right)  + 2 \rho_{mp}  \left( z_{n} z_{t} \right) \left(  \rho_{ls} + \delta_{ls} \right) \\  + 2  \left( z_{p}^{*} z_{m} \right)  \kappa_{nt}  \left(  \rho_{ls} + \delta_{ls} \right) + \kappa_{ml}  \kappa_{nt} \kappa_{ps}^{*}  + 2 \kappa_{ml}  \left( z_{n} z_{t} \right) \kappa_{ps}^{*} +\kappa_{ml}  \kappa_{nt}  \left( z_{p}^{*} z_{s}^{*} \right) \end{array} \right\}
\end{eqnarray}

Formal integration of the equations of motion (\ref{CondKin})-(\ref{QPKin}) (taking into account their suppressed `free' evolution in the HFB basis) and substitution into equations (\ref{CondProp})-(\ref{QPProp}) generates all second order collisional processes, as discussed in more detail in the next section. Before proceeding, we give at this point the most compact form of our equations to second order as
\begin{eqnarray}
i  \left( \frac{dR_{c}}{dt} \right) = \left( \frac{ \partial \hat{E} }{ \partial R_{c} } \right) R_{c}
\end{eqnarray}
\begin{eqnarray}
i \left( \frac{dR_{e}}{dt} \right)_{ji} = \left[ \sum_{l} \left( \frac{ \partial \hat{E} }{ \partial (R_{e})_{lj} } \right) \left( R_{e} \right)_{li} -i \left( I_{ji} \right) \right] - h.c.
\end{eqnarray}
where the matrix $I$ is defined as
\begin{eqnarray}
I = \left( \begin{array}{c} I_{\rho} \\ I_{\kappa}^{*} \end{array} \begin{array}{c} I_{\kappa} \\ I_{\rho}^{*} \end{array} \right)
\end{eqnarray}

\section{Link to other kinetic theories}

Having discussed the extent to which a variational methodology can be useful, let us now show explicitly how our equations reproduce those of Walser et al. \cite{Walser_QK}, as written in generalised matrix form in their follow-up paper \cite{Walser_Sim}. By thinking in terms of forward and backward scattering rates incorporating terms $I_{\rho}$ and $I_{\kappa}$, the equations of motion for the non-equilibrium matrix $K$  can be straightforwardly re-expressed as

\begin{eqnarray}
i \frac{dK}{dt} = & &\left( \begin{array}{c} \left( \Gamma_{\rho}- I_{\rho} \right) \\ - I_{\kappa}^{*}  \end{array} \begin{array}{c} \left( \Gamma_{\kappa} + I_{\kappa} \right)   \\  I_{\rho}^{*}  \end{array} \right)  \left( \begin{array}{c}  \rho \\  \kappa^{*} \end{array} \begin{array}{c} \kappa \\  \left( \rho^{*} + {\bf \un} \right) \end{array} \right) \nonumber \\  
& - & \left( \begin{array}{c} - I_{\rho}  \\ - \left[ \Gamma_{\kappa} + I_{\kappa} \right]^{*}  \end{array} \begin{array}{c}  I_{\kappa}   \\ - \left[ \Gamma_{\rho} - I_{\rho} \right]^{*}  \end{array} \right)  
\left( \begin{array}{c}  \left( \rho + \un \right) \\  \kappa^{*} \end{array} \begin{array}{c} \kappa \\ \rho^{*} \end{array} \right) + h.c.
\end{eqnarray}
Following the notation of Walser et al. \cite{Walser_Sim}, we now define forward and backward collision operators as
\begin{eqnarray}
\Gamma^{<} = \left( \begin{array}{c} - I_{\rho}  \\ - \left[ \Gamma_{\kappa} + I_{\kappa} \right]^{*}  \end{array} \begin{array}{c}  I_{\kappa}   \\ - \left[ \Gamma_{\rho} - I_{\rho} \right]^{*}  \end{array} \right)  
\end{eqnarray}
and $\Gamma^{>} = - \sigma_{1} \left( \Gamma^{<} \right)^{*}  \sigma_{1}$. The corresponding generalised quasiparticle self-energies are $R_{e}^{>} = R_{e}$ as defined in equation (23) and $R_{e}^{<} = \sigma_{1} \left( R_{e}^{>} \right)^{*}  \sigma_{1}$. We thus obtain 

\begin{eqnarray}
i \frac{d K}{dt} = - \left[ \left( \Gamma^{<} R_{e}^{<} - \Gamma^{>} R_{e}^{>} \right) + h.c. \right]
\end{eqnarray}
yielding for the quasiparticle propagator the following second order expression
\begin{eqnarray}
\frac{d R_{e}^{>}}{dt} = -i H_{e} R_{e}^{>} + \left( \Gamma^{<} R_{e}^{<} - \Gamma^{>} R_{e}^{>} \right) + h.c.
\end{eqnarray}
Careful examination of the above equation shows clearly that our approach generates exactly the same forward and backward collision operators for the non-condensate propagator, as discussed in \cite{Walser_Sim}.

Although our methodology generates the forward and backward collision operators for the non-condensate propagator in a rather straightforward manner, the situation becomes somewhat more complicated when considering the evolution of the condensed component. The reason is that our methodology does not yield such expressions directly, but instead it provides us with the {\em sum} of forward and backward rates, via
\begin{eqnarray}
\fl \left( \frac{dR_{c}}{dt} \right) = - i \left( \frac{ \partial \hat{E} }{ \partial R_{c} } \right) R_{c} =  - i \left( \frac{ \partial \hat{E}_{HFB}^{(1)} }{ \partial R_{c} } +   \frac{ \partial \hat{E}^{(2)} }{ \partial R_{c} } \right) R_{c} =  - i H_{c} R_{c} + Y R_{c}
\end{eqnarray}
where
\begin{eqnarray}
Y = \left( \begin{array}{c} - \Gamma_{z} \\ - \left[ \Gamma_{z^{*}} \right]^{*}  \end{array} \begin{array}{c} - \Gamma_{z^{*}}  \\ - \left[ \Gamma_{z} \right]^{*}  \end{array} \right)
\end{eqnarray}
thus leaving their particular definition somewhat arbitrary. A natural definition corresponds to identifying these rates in terms of positive and negative contributions of $\Gamma_{z^{(*)}}$ (denoted respectively by $\pm ve[\Gamma_{z^{(*)}}]$), thus defining the forward and backward scattering rates for the condensate propagator as
\begin{eqnarray}
Y^{<} = \left( \begin{array}{c} -ve \left[ \Gamma_{z} \right]  \\  - \left( +ve \left[ \Gamma_{z^{*}} \right] \right)^{*}   \end{array} \begin{array}{c}  -ve \left[ \Gamma_{z^{*}} \right]   \\  - \left( +ve \left[ \Gamma_{z} \right] \right)^{*}   \end{array} \right)  
\end{eqnarray}
and 
\begin{eqnarray}
Y^{>}  = - \sigma_{1} \left( Y^{<} \right)^{*}  \sigma_{1} =  \left( \begin{array}{c} +ve \left[ \Gamma_{z} \right]  \\ - \left(  -ve \left[ \Gamma_{z^{*}} \right] \right)^{*}   \end{array} \begin{array}{c}  +ve \left[ \Gamma_{z^{*}} \right]   \\  - \left( -ve \left[ \Gamma_{z} \right] \right)^{*}   \end{array} \right)  
\end{eqnarray}
which clearly satisfy $Y = \left( Y^{<} - Y^{>} \right)$.

In comparing our final expressions to those of Walser et al. \cite{Walser_QK,Walser_Sim}, we note that our combined evolution equations agree entirely with the {\em first} formulation of the theory of Walser et al. \cite{Walser_QK}. In their subsequent work \cite{Walser_Sim}, where they have explicitly identified forward and backward rates as described in this section, their expressions for {\em condensate} rates differ from those we have given above in that their corresponding expression for $\Gamma_{z^{*}}$ of equation (\ref{Gammaz}) is
\begin{eqnarray}
\fl \left( \Gamma_{z^{*}} \right)_{iq} = 2 \sum_{smn} \sum_{pqlt} V_{jsmn} \tilde{V}_{pqlt} \left\{ \begin{array}{l} 2 \left[ \left( \rho_{mp} + \delta_{mp} \right) \kappa_{nt} \rho_{ls} -  \rho_{mp} \kappa_{nt} \left( \rho_{ls} + \delta_{ls} \right) \right] \\ + \left[ \kappa_{ml} \kappa_{nt} \kappa_{ps}^{*} - \kappa_{ml} \kappa_{nt} \kappa_{ps}^{*} \right] \end{array}  \right\} 
\end{eqnarray} 
with the second line of the above expression, which {\em identically cancels itself}, being generated {\em additionally} to what appears in our expression of equation (\ref{Gammaz}). The appearance of such a contribution is physically appealing, since it implies that the collision processes  occuring in condensate and non-condensate are of the same basic structure \cite{Walser_Sim}, indicating that processes $\Gamma^{>}$ can be generated from $Y^{>}$ by functional differentiation, as done in Green functions' techniques \cite{Martin,Kadanoff}. Although the net contribution of such a term is zero, its presence modifies the physical expressions of forward and backward collision integrals, i.e. it adds a particular collisional diagram to each. In our treatment, it is clear that such an extra `self-cancelling' contribution cannot be generated in the first place. This is because the adiabatic elimination procedure yields for the second order expression for $z$ terms proportional, at most, to correlations of five operators, whereas these additional contributions would require correlations of seven single-particle operators, and hence a treatment of three-particle interactions \cite{Gora}, which extends beyond the binary hamiltonian employed here. It is therefore somewhat perplexing that such terms are found in the second formulation of the theory of Walser et al. \cite{Walser_Sim}. However, if one is looking for a theory which satisfies such symmetries as mentioned above, due to specific physical considerations, we can indeed {\em add and subtract} these (or other similar) contributions to the forward and backward {\em condensate} collision integrals, obtaining {\em exactly} the same contributions as those discussed by Walser et al. in \cite{Walser_Sim} and by Wachter et al. \cite{Wachter,Wachter_Thesis}. Such treatment will, however, not be a rigorous derivation, since its last stage will rely on additional physical arguments (e.g. gapless or conserving \cite{Martin,Milena_Book,Griffin_Gap}) arising from alternative approaches.

We can now summarize the entire findings of our perturbative HFB treatment in the following systems of equations
\begin{eqnarray}
 \left\{ \begin{array}{c} i \frac{d R_{c}}{dt} =  H_{c} R_{c} + J \\
i \frac{d R_{e}^{>}}{dt} =  H_{e} R_{e}^{>} - R_{e}^{>} H_{e}^{\dag} + K \end{array} \right\}
\end{eqnarray}

\begin{eqnarray}
\left\{ \begin{array}{c} i \frac{d J}{dt} = - \left( Y^{<}  - Y^{>} \right) R_{c} \\  i  \frac{d K}{dt} = - \left[ \left( \Gamma^{<} R_{e}^{<} - \Gamma^{>} R_{e}^{>} \right) + h.c. \right] \end{array} \right\}
\end{eqnarray}
Combining these results yields for the propagators of the system the following second order evolution
\begin{eqnarray}
\left\{ \begin{array}{c} \frac{d R_{c}}{dt} =  - i H_{c} R_{c} + \left( Y^{<}  - Y^{>} \right) R_{c} \\ 
\frac{d R_{e}^{>}}{dt} =  -i H_{e} R_{e}^{>} +  \left( \Gamma^{<} R_{e}^{<} - \Gamma^{>} R_{e}^{>} \right) + h.c.  \end{array} \right\}
\end{eqnarray}

\section{Conclusions}

By applying second order perturbation theory from an HFB basis and making no further approximations on the binary interaction hamiltonian, we have generalised an earlier quantum kinetic approach to Bose-Einstein condensation presented by the author and Burnett elsewhere \cite{Prouk_NIST,Prouk_Thesis}, by treating in a self-consistent manner the dynamics of both condensed and non-condensed components of the system. This involves a detailed consideration of the evolution of correlations of up to four shifted single-particle operators. Our methodology can be summarised as follows: Firstly we obtain the exact equations of motion for the condensate and non-condensate propagators in the binary interaction hamiltonian. These depend on matrices expressed in terms of higher order correlations, which can be approximately treated by separating off their `equilibrium' and `non-equilibrium' contributions (somewhat analogously to \cite{Giorgini,Rusch}). The `equilibrium' contributions are decomposed into lower order averages in terms of the HFB order parameters (thus yielding the usual first order evolution), whereas the non-equilibrium matrices  contain {\em all} contributions of order higher than those obtained  from the HFB energy functional. In this paper, we have focused on the lowest order expression for the evolution of such non-equilibrium matrices. Substitution of these results into the exact equations of motion for the condensate and non-condensate propagators yields their respective evolution to second order in the effective interatomic potentials and these are shown to be identical to those of Walser et al. \cite{Walser_QK,Walser_Sim}. Such an equivalence was of course anticipated, since they are both based on essentially the same underlying assumption of slowly-varying quantities. In the work of Walser et al. \cite{Walser_QK}, these quantities correspond to the `master' variables determining the evolution of the gas on a coarse-grained timescale, whereas in the treatment presented here, these quantities are implicit in our choice of the HFB hamiltonian as the unperturbed hamiltonian for applying perturbation theory. Our approach has provided the {\em exact} form of the energy functional (equations (\ref{EFTotal})-(\ref{EF2})), which readily enables us to extend this treatement to higher orders. This may be useful, for example, to test a key underlying assumption of all such second order theories  for dilute, weakly-interacting Bose-condensed gases, that the higher order terms (in the perturbative expansion) are significantly smaller. This should be valid sufficiently far from the critical region, when $(kT/nU_{o})(\sqrt{n a^{3}}) \ll 1$ \cite{Popov,Gora} where $a$ is the s-wave scattering length, $n$ the condensate density and $U_{o}$ the effective two-body interatomic potential (and can thus be justified by the fact that the problem has been formulated in terms of an unperturbed {\em quasiparticle} basis and that the interactions have been treated by an {\em effective resummed} bare t-matrix).

Furthermore, our work is based on the same idea as the second order theory of excitations developed by Morgan \cite{Morgan} and recently applied numerically by the Oxford group \cite{Morgan_Sim}, thus showing the link between these two approaches.  We should also point out that our approach is formally connected to the detailed kinetic treatment of Zaremba-Nikuni-Griffin \cite{ZNG} (whose analysis focuses on the hydrodynamic regime). Moreover, in a recent paper \cite{Wachter}, the theory of Walser et al. \cite{Walser_QK,Walser_Sim} has been shown to be equivalent to the Green's function methodology originally developed by Kadanoff and Baym \cite{Kadanoff} and recently applied to the study of trapped condensates by Imamovic-Tomasovic and Griffin \cite{Milena_Book,Milena_HFB}. This latter approach is in turn formally related to the Schwinger-Keldysh method for treating non-equilibrium many-body systems, which has been employed by Stoof \cite{Stoof_FP}. We stress that the work we have presented in this paper includes all relevant quasiparticle effects, despite being expressed in terms of single-particle operators; to generalise this, one could explicitly transform to quasiparticle operators. Carrying out such a transformation within the framework of the kinetic theory of Walser et al. \cite{Walser_QK,Walser_Sim}, Wachter has shown \cite{Wachter_Thesis} that the collisional integrals simplify enormously, acquiring the usual form of Boltzmann-like factors, thus enabling a direct physical interpretation (which is rather concealed in the lengthy expressions in terms of single-particle operators). Similar work has been performed by Imamovic-Tomasovic and Griffin who recently described such a quasiparticle kinetic equation \cite{Milena_QP}. Here we should further point out that the static theory of Morgan \cite{Morgan} was formulated explicitly in terms of quasiparticles, and so where the related linear response discussions of Rusch et al. \cite{Rusch} and Giorgini \cite{Giorgini}. The links of the current approach independently to those of Walser et al. (and hence Imamovic-Tomasovic and Griffin) and Morgan suggest that these time-dependent theories may correspond to some form of time-dependent generalisation of the gapless excitation theory of Morgan, although a more direct link remains to be established. Based on these considerations we hence believe that this paper represents yet another contribution towards the long-sought goal of a universally accepted kinetic theory for dilute, weakly-interacting systems exhibiting Bose-Einstein condensation.

\ack

I am grateful to Peter Lambropoulos for detailed discussions. I would also like to thank Reinhold Walser for in-depth discussions during the First International Workshop on the Theory of Quantum Gases and Quantum Coherence held in Salerno, Italy, and Keith Burnett for various general discussions on quantum kinetic theory during the past few years. This work was partially supported by the ULF (Contract No. ERB-FMGECT 950021).

\appendix \section{Equations of motion for multiple shifted single-particle operator correlations}

In this appendix we give expressions for the equations of motion of correlations of up to four shifted single-particle operators. Our expressions are only explicit for the effect of the reduced hamiltonian $\left( \hat{H}_{3}^{'} +  \hat{H}_{4}^{'} \right)$, with the effect of the HFB hamiltonian $\left( H_{0} + \hat{H}_{1}^{'} +  \hat{H}_{2}^{'} \right)$  being implicit in the renormalised eigenenergies $\tilde{\omega}_{i}$ appearing below (and also in the approximately energy-conserving matrix elements $\tilde{V}_{pqlt}$ of equation (\ref{Matrix})). Explicit expressions in terms of the {\em entire} hamiltonian of the system $\hat{H}$ of equation (\ref{Ham}) (including also the ones needed to derive the first order HFB-basis contributions to the coupled dynamics, i.e. first terms in each of equations (\ref{CondProp})-(\ref{QPProp})) can be found in \cite{Prouk_NIST,Prouk_Thesis}. We start by noting that the HFB quantities $z, \rho, \kappa$ will, to lowest order in the potential, have no contribution in the hamiltonian  $\left( \hat{H}_{3}^{'} +  \hat{H}_{4}^{'} \right)$, since the quantities found on their right-hand-sides acquire no equilibrium mean value within the HFB approximation. In particular, we find (making use of symmetries wite respect to summed indices)
\begin{equation}
i \frac{d}{dt} \left( z_{i} \right) = \tilde{\omega}_{i} z_{i} + \sum_{smn} V_{ismn} \lambda_{smn} \label{z}
\end{equation}
\begin{eqnarray} 
i \frac{d}{dt} \left( \rho_{ji} \right) & = \left( \tilde{\omega}_{j} -  \tilde{\omega}_{i} \right) \rho_{ji} +  \sum_{smn} V_{jsmn} \left[ \sigma_{ismn}  +\left( 2 \lambda_{mis}^{*} z_{n} + \lambda_{imn} z_{s}^{*} \right) \right] \nonumber \\
& - \sum_{smn} V_{mnsi} \left[ \sigma_{mnsj} +\left( 2 \lambda_{mjs} z_{n}^{*} + \lambda_{jmn}^{*} z_{s} \right) \right] \label{rho}
\end{eqnarray}
\begin{eqnarray} 
i \frac{d}{dt} \left( \kappa_{ji} \right) & = \left( \tilde{\omega}_{i} +  \tilde{\omega}_{j} \right) \kappa_{ji} +  \sum_{smn} V_{jsmn} \left[ \xi_{simn} + \left( 2 \lambda_{sim} z_{n} + \gamma_{imn} z_{s}^{*} \right) \right] \nonumber \\
& + \sum_{smn} V_{ismn} \left[ \xi_{sjmn} + \left( 2 \lambda_{sjm} z_{n} + \gamma_{jmn} z_{s}^{*} \right) \right] \label{kappa}
\end{eqnarray}
where the entire first order evolution of condensate / non-condensate propagators has now been summed up in the renormalised HFB eigenenergies $\tilde{\omega}_{i}$.

For the non-HFB quantities, we find the following evolution
\begin{eqnarray} 
i \frac{d}{dt} \left( \lambda_{smn} \right) & = \left( \tilde{\omega}_{m} + \tilde{\omega}_{n} - \tilde{\omega}_{s} \right) \lambda_{smn} + \sum_{lt} V_{mnlt} \left( 2 \rho_{ls} z_{t} \right) \nonumber \\
& +  \sum_{plt} V_{pnlt} \left[ 2 \rho_{mp} \rho_{ls} z_{t} + 2 \kappa_{ml} \kappa_{ps}^{*} z_{t} + 2 \rho_{ls} \kappa_{mt} z_{p}^{*} \right] \nonumber \\
& +  \sum_{plt} V_{pmlt} \left[ 2 \rho_{np} \rho_{ls} z_{t} + 2 \kappa_{nl} \kappa_{ps}^{*} z_{t} + 2 \rho_{ls} \kappa_{nt} z_{p}^{*} \right] \nonumber \\
& - \sum_{pql} V_{pqls} \left[2 \rho_{mp} \rho_{nq} z_{l} + 2 \left( \rho_{mp} \kappa_{nl} + \rho_{np} \kappa_{ml} \right) z_{q}^{*} \right] \label{lambda}
\end{eqnarray}

\begin{eqnarray} 
i  \frac{d}{dt} \left( \gamma_{imn} \right) & = \left( \tilde{\omega}_{i} + \tilde{\omega}_{m} + \tilde{\omega}_{n} \right) \gamma_{imn} \nonumber \\ 
& + 2 \sum_{lt} \left[ V_{mnlt} \left( \kappa_{il}  z_{t} \right) + V_{inlt} \left( \kappa_{ml}  z_{t} \right) +   V_{imlt} \left( \kappa_{nl}  z_{t} \right) \right] \nonumber \\
& + 2 \sum_{plt} V_{pnlt} \left[ \rho_{ip} \kappa_{ml} z_{t} + \rho_{mp} \kappa_{il} z_{t} + \kappa_{il} \kappa_{mt} z_{p}^{*} \right] \nonumber \\
& + 2 \sum_{plt} V_{pmlt} \left[ \rho_{ip} \kappa_{nl} z_{t} + \rho_{np} \kappa_{il} z_{t} + \kappa_{il} \kappa_{nt} z_{p}^{*} \right] \nonumber \\
& + 2 \sum_{plt} V_{pilt} \left[ \rho_{mp} \kappa_{nl} z_{t} + \rho_{np} \kappa_{ml} z_{t} + 2 \kappa_{ml} \kappa_{nt} z_{p}^{*} \right] \label{gamma}
\end{eqnarray}

\begin{eqnarray} 
i  \frac{d}{dt} \left( \sigma_{ismn} \right) & = \left( \tilde{\omega}_{m} + \tilde{\omega}_{n} - \tilde{\omega}_{i} - \tilde{\omega}_{s} \right) \sigma_{ismn} \nonumber \\
& + \sum_{lt} V_{mnlt} \left( 2 \rho_{li} \rho_{ts} \right) - \sum_{pq} V_{pqis} \left( 2 \rho_{mp} \rho_{nq} \right) \nonumber \\
& + 2 \sum_{plt} V_{pnlt} \left( \rho_{mp} \rho_{li} \rho_{ts} + \rho_{ls} \kappa_{mt} \kappa_{pi}^{*} +  \rho_{li} \kappa_{mt} \kappa_{ps}^{*}\right) \nonumber \\
& + 2 \sum_{plt} V_{pmlt} \left( \rho_{np} \rho_{li} \rho_{ts} + \rho_{ls} \kappa_{nt} \kappa_{pi}^{*} +  \rho_{li} \kappa_{nt} \kappa_{ps}^{*}\right) \nonumber \\
& - 2 \sum_{pql} V_{pqli} \left( \rho_{mq} \rho_{np} \rho_{ls} + 2 \rho_{mp} \kappa_{nl} \kappa_{qs}^{*} \right)  \nonumber \\ 
& - 2 \sum_{pql} V_{pqls} \left( \rho_{mq} \rho_{np} \rho_{li} + 2 \rho_{mp} \kappa_{nl} \kappa_{qi}^{*} \right) \label{sigma}
\end{eqnarray}

\begin{eqnarray} 
i \frac{d}{dt} \left( \xi_{simn} \right) & = \left( \tilde{\omega}_{i} + \tilde{\omega}_{m} + \tilde{\omega}_{n} - \tilde{\omega}_{s} \right) \xi_{simn} \nonumber \\
& + \sum_{lt} \left[ V_{mnlt} \left( 2 \rho_{ls} \kappa_{it} \right) +  V_{inlt}   \left( 2 \rho_{ls} \kappa_{mt} \right) +  V_{imlt} \left( 2 \rho_{ls} \kappa_{nt} \right) \right] \nonumber \\
& + 2 \sum_{plt}  V_{pnlt} \left[  \kappa_{ml} \kappa_{it} \kappa_{ps}^{*} +  \rho_{mp} \rho_{ls} \kappa_{it} + \rho_{ip} \rho_{ls} \kappa_{mt} \right] \nonumber \\
& + 2 \sum_{plt} V_{pmlt} \left[ \kappa_{nl} \kappa_{it} \kappa_{ps}^{*} +  \rho_{np} \rho_{ls} \kappa_{it} + \rho_{ip} \rho_{ls} \kappa_{nt} \right]  \nonumber \\
& + 2  \sum_{plt} V_{pilt} \left[  \kappa_{nl} \kappa_{mt} \kappa_{ps}^{*} +  \rho_{np} \rho_{ls} \kappa_{mt} + \rho_{mp} \rho_{ls} \kappa_{nt} \right] \nonumber \\
& - 2 \sum_{pql}  V_{pqls} \left[   \rho_{mp} \rho_{nq} \kappa_{li} + \rho_{mp} \rho_{iq} \kappa_{nl} +\rho_{ip} \rho_{nq} \kappa_{ml}  \right] \label{xi}
\end{eqnarray}

We stress that equations (\ref{lambda})-(\ref{xi}) given above are not exact, even within our consistent decoupling approximations, because we have systematically ignored from their right hand sides contributions which depend on `non-equilibrium' correlations of more than two shifted single-particle operators (i.e. $\lambda$, $\gamma$, $\sigma$, or $\xi$). Such contributions will ultimately yield the evolution of the generalised condensate / quasiparticle propagators beyond second order in the potential, in the same manner that equations (\ref{z})-(\ref{kappa}) will also yield contributions to third or higher orders. We believe this is fully equivalent to the statement made in Appendix A of Walser et al. \cite{Walser_QK}, that terms leading to higher order contributions are consistently neglected. We could, of course, have very straightforwardly included such terms in the equations of motion  (\ref{lambda})-(\ref{xi}). However, since we do not deal with such terms any further, this would only introduce unnecessary complexity.

\section{Decoupling approximations}

As mentioned in the text, correlations of even numbers of shifted single-particle operators,  can be thought of as consisting of two contributions \cite{Prouk_Thesis}, one part being due to the mean value at equilibrium when decomposed by Wick's theorem into products of lower averages (here referring to binary averages) and the remaining term yielding the rate of change of this quantity from its equilibrium value. For example
$ \langle \hat{c}_{r}^{\dag} \hat{c}_{s}^{\dag} \hat{c}_{m} \hat{c}_{n} \rangle = \langle \hat{c}_{r}^{\dag} \hat{c}_{s}^{\dag} \hat{c}_{m} \hat{c}_{n} \rangle^{0} +\sigma_{rsmn}$
with the `non-equilibrium' quantities (here $\sigma$) treated perturbatively, by means of their respective equations of motion given in Appendix A.

Since the HFB basis, by definition, does not allow for nonzero equilibrium values of triplet correlations, we note that, by analogy, {\em any} correlations of {\em odd} products of shifted single-particle operators will vanish to lowest order (since their final decomposed forms will always depend on such triplet products as discussed in \cite{Prouk_NIST,Prouk_Thesis}) and will yield non-vanishing contributions only at the next order in the potential, via their respective equations of motions discussed in Appendix A.
The decoupling approximations we have employed for the `equilibrium' quantities are as follows
\begin{equation}
\langle \hat{c}_{r}^{\dag} \hat{c}_{s}^{\dag} \hat{c}_{m} \hat{c}_{n} \rangle^{0} = \rho_{mr} \rho_{ns} + \rho_{nr} \rho_{ms} + \kappa_{mn} \kappa_{rs}^{*}
\end{equation}
At equilibrium, this would arise directly from Wick's theorem, which is the main motivation for such a decoupling approximation here. The remaining decorrelations are performed analogously, namely
\begin{equation}
\langle \hat{c}_{p}^{\dag} \hat{c}_{q} \hat{c}_{l} \hat{c}_{t} \rangle =\rho_{qp} \kappa_{lt} +\rho_{lp} \kappa_{qt} + \rho_{tp} \kappa_{ql} 
\end{equation}
\begin{equation}
\langle \hat{c}_{p} \hat{c}_{q} \hat{c}_{l} \hat{c}_{t} \rangle = \kappa_{pq}  \kappa_{lt} + \kappa_{pl}  \kappa_{qt} + \kappa_{pt}  \kappa_{ql} 
\end{equation}
\begin{eqnarray}
\fl \six{p}{r}{s}{q}{l}{t} \nonumber \\
\fl = \left\{ \begin{array}{l} \rho_{qp} \left( \rho_{lr} \rho_{ts}  + \rho_{tr} \rho_{ls} \right) + \rho_{lp} \left( \rho_{qr} \rho_{ts} + \rho_{tr} \rho_{qs} \right) +   \rho_{tp} \left( \rho_{qr} \rho_{ls} + \rho_{lr} \rho_{qs} \right) \\
+ \kappa_{ql} \left( \kappa_{pr}^{*} \rho_{ts} + \kappa_{ps}^{*} \rho_{tr} + \kappa_{rs}^{*} \rho_{tp} \right) + \kappa_{qt} \left( \kappa_{pr}^{*} \rho_{ls} + \kappa_{ps}^{*} \rho_{lr} + \kappa_{rs}^{*} \rho_{lp} \right) \\
+ \kappa_{lt} \left( \kappa_{pr}^{*} \rho_{qs} + \kappa_{ps}^{*} \rho_{qr} + \kappa_{rs}^{*} \rho_{qp} \right) \end{array} \right\}
\end{eqnarray}

\begin{eqnarray} 
\fl \langle \hat{c}_{p}^{\dag} \hat{c}_{q}^{\dag} \hat{c}_{m} \hat{c}_{r} \hat{c}_{l} \hat{c}_{t} \rangle \nonumber \\
\fl = \left\{ \begin{array}{l}  \rho_{mp} \left( \rho_{rq} \kappa_{lt} + \rho_{lq} \kappa_{rt} + \rho_{tq} \kappa_{rl} \right)  + \rho_{rp} \left( \rho_{mq} \kappa_{lt} + \rho_{lq} \kappa_{mt} + \rho_{tq} \kappa_{ml} \right) \\
+ \rho_{lp} \left( \rho_{rq} \kappa_{mt} + \rho_{mq} \kappa_{rt} + \rho_{tq} \kappa_{mr} \right)  + \rho_{tp} \left( \rho_{rq} \kappa_{lm} + \rho_{mq} \kappa_{rl} + \rho_{lq} \kappa_{mr} \right) \\ + \kappa_{pq}^{*} \left( \kappa_{rm} \kappa_{lt} + \kappa_{lm} \kappa_{rt} + \kappa_{tm} \kappa_{rl} \right) \end{array} \right\}
\end{eqnarray}


\section*{References}

\begin{thebibliography}{99}


\bibitem{Martin} Hohenberg P C and Martin P C 1965 {\it Ann. Phys. (NY)} {\bf 34} 291 [reprinted 2000 {\it Ann. Phys. (NY)} {\bf 281} 636]\nonum Martin P C 1961 {\it J. Math. Phys.} {\bf 4} 208

\bibitem{Kadanoff} Kane J W and Kadanoff L P 1965 {\it J. Math. Phys.} {\bf 6} 1902\nonum L.P. Kadanoff and G. Baym, {\it Quantum Statistical Mechanics} (W.A. Benjamin, Inc., New York, 1962).

\bibitem{Stoof_FP}  Stoof H T C 1999 {\it J. Low Temp. Phys.} {\bf 114} 11

\bibitem{QKV} Gardiner C W and Zoller P 2000 {\it Phys. Rev. A} {\bf 61} 033601

\bibitem{MIT_Growth} Miesner H-J, Stamper-Kurn D M, Andrews M R, Durfee D S, Inouye S and Ketterle W 1998 {\it Science} {\bf 279} 1005 

\bibitem{Stoof_Growth} Bijlsma M J, Zaremba E and Stoof H T C 2000 {\it Phys. Rev. A} {\bf 62} 063609 

\bibitem{QK_Growth} Lee M D and Gardiner C W 2000 {\it Phys. Rev. A} {\bf 62} 033606\nonum  Davis M J, Gardiner C W and Ballagh R J 2000 {\it Phys. Rev. A} {\bf 62} 063608

\bibitem{Kagan_QK} Svistunov B V 1991 {\it J. Moscow Phys. Soc.} {\bf 1} 373\nonum Kagan Y, Svistunov B V and Shlyapnikov G V 1992 {\it Zh. Eksp. Teor. Fiz.} {\bf 101} 528 [{\it 1992 Sov. Phys. JETP} {\bf 75} 387]

\bibitem{Castin_U1} Castin Y and Dum R 1998 {\it Phys Rev A} {\bf 57} 3008

\bibitem{Kirkpatrick} Kirkpatrick T R and Dorfmann J R 1985 {\it J. Low Temp. Phys.} {\bf 58} 301; {\it ibid} 399

\bibitem{Prouk_NIST} Proukakis N P and Burnett K 1996 {\it J. Res. Natl. Inst. Stand. Technol.}  {\bf 101} 457 

\bibitem{ZNG} Zaremba E, Nikuni T and Griffin A 1999 {\it J. Low Temp. Phys.} {\bf 116} 277

\bibitem{Walser_QK} Walser R, Williams J, Cooper J and Holland M 1999 {\it Phys. Rev. A} {\bf 59} 3878 

\bibitem{Walser_Sim} Walser R, Cooper J and Holland M 2000  {\it Phys. Rev. A} {\bf 63} 013607

\bibitem{Wachter} Wachter J, Walser R, Cooper J and Holland M 2001 {\it preprint (cond-mat/0105181)}

\bibitem{Milena_Book} Imamovic-Tomasovic M and Griffin A 2000 {\it Progress in Nonequilibrium Green's Functions}, M. Bonitz (ed.) (World Scientific: Singapore) 

\bibitem{Milena_HFB} Imamovic-Tomasovic M and Griffin A 1999 {\it Phys. Rev. A} {\bf 60} 494 

\bibitem{Beliaev}  Beliaev S T 1958 {\it Sov. Phys. JETP} {\bf 34} 289; {\it ibid.} 299 

\bibitem{Popov} Popov V N 1965 {\it Sov. Phys. JETP} {\bf 20} 1185\nonum Popov V N 1983 {\it Functional Intergrals in Quantum Field Theory and Statistical Physics} (Reidel: Dodrecht)

\bibitem{Gora} Fedichev P O and Shlyapnikov G V 1998 {\it Phys. Rev. A} {\bf 58} 3146

\bibitem{Prouk_T_Matrix} Proukakis N P, Burnett K and Stoof H T C 1998 {\it Phys. Rev. A} {\bf 57} 1230

\bibitem{Prouk_1D} Proukakis N P, Morgan S A, Choi S and Burnett K 1998 {\it Phys. Rev. A} {\bf 58} 2435 

\bibitem{Bijlsma_Var} Bijlsma M and Stoof H T C 1997 {\it Phys. Rev. A} {\bf 55} 498 

\bibitem{Shi} Shi H and Griffin A 1998 {\it Phys. Rep.} {\bf 304} Nos. 1-2, 1 

\bibitem{Morgan} Morgan S A 2000 {\it J. Phys. B: At. Mol. Opt. Phys.} {\bf 33} 3847 

\bibitem{Gavoret} Gavoret J and Nozieres P 1964 {\it Ann. Phys.} {\bf 28} 349\nonum Nepomnyashchii Y A and Nepomnyashchii A A 1979 {\it Sov. Phys. JETP} {\bf 48} 493 [1978 {\it Zh. Eksp. Teor. Fiz.} {\bf 75} 976]

\bibitem{LHY} Lee T D, Huang K and Yang C N 1957 {\it  Phys. Rev.} {\bf 106} 1135

\bibitem{Gap} Girardeau M and Arnowitt A 1959 {\it Phys. Rev.} {\bf 113} 755\nonum Wentzel G 1960 {\it Phys. Rev.} {\bf 120} 1572 

\bibitem{Prouk_Thesis} Proukakis N P 1997 {\it DPhil Thesis} (University of Oxford)

\bibitem{Huang} Huang K 1987 {\it Statistical Mechanics} (2nd ed., John Wiley and Sons)

\bibitem{Prouk_JPhysB} Hutchinson D A W, Burnett K, Dodd R J, Morgan S A, Rusch M, Zaremba E, Proukakis N P, Edwards M and Clark C W 2000 {\it J. Phys. B: At. Mol. Opt. Phys.} {\bf 33} 3825 

\bibitem{Blaizot} Blaizot J P and Ripka G 1986 {\it Quantum Theory of Finite Systems} (MIT Press)

\bibitem{Griffin_Gap} Griffin A 1996 {\it  Phys. Rev. B} {\bf  53} 9341 

\bibitem{Giorgini} Giorgini S 1998 {\it Phys. Rev. A} {\bf 57} 2949\nonum Giorgini S 2000 {\it Phys. Rev. A} {\bf 61} 063615 

\bibitem{Bogoliubov} Bogoliubov N N 1947 {\it J. Phys. U.S.S.R.} {\bf 11} 23 

\bibitem{Takano} Takano F 1961 {\it Phys. Rev.} {\bf 123} 699 


\bibitem{Milena_QP} Imamovic-Tomasovic M and Griffin A 2001 {\it J. Low Temp. Phys.} {\bf 122} 617 

\bibitem{Wachter_Thesis} Wachter J 2000 {\it  Master's Thesis}  (Univ. of Boulder, Colorado)

\bibitem{Rusch} Rusch M and Burnett K 1999 {\it Phys. Rev. A} {\bf 59} 3851

\bibitem{Morozov} Morozov V G and  R\"{o}pke G 1999 {\it Ann. Phys.} {\bf 278} 127

\bibitem{Morgan_Sim} Rusch M, Morgan S A, Hutchinson D A W and Burnett K 2000 {\it Phys. Rev. Lett.} {\bf 85} 4844


\end{thebibliography}
\end{document}